\documentclass[12pt,a4paper]{article}
\title{Comments on the global constraints\\
in light-cone string and membrane theories}
\author{{\sc Shozo Uehara}\footnote{e-mail:
uehara@eken.phys.nagoya-u.ac.jp}~ and
{\sc Satoshi Yamada}\footnote{e-mail:
yamada@eken.phys.nagoya-u.ac.jp}\vspace{4mm}\\
{\sl Department of Physics, Nagoya University}\\
{\sl Chikusa-ku, Nagoya 464-8602, Japan}}
\date{}

\makeatletter

\@addtoreset{equation}{section}
\renewcommand{\thefigure}{\@arabic\c@figure}
\makeatother

\newcommand{\nn}{\nonumber\\}
\newcommand{\ptau}{\partial_\tau}
\newcommand{\psig}{\partial_\sigma}
\newcommand{\prho}{\partial_\rho}
\newcommand{\brs}{\delta_B}
\newcommand{\To}{\Rightarrow}
\newcommand{\p}{\partial}
\newcommand{\ket}[1]{|#1\rangle}

\begin{document}
\maketitle
\vspace{-80mm}
\begin{flushright}
DPNU-02-41\\
hep-th/0212048\\
December 2002
\end{flushright}
\vspace{57mm}

%%%%%%%%%%%%%%%%%%%%%%%%%%%%%%%%%
\begin{abstract}
In the light-cone closed string and toroidal membrane theories,
we associate the global constraints with gauge symmetries.
In the closed string case, we show that the physical states defined by
the BRS charge satisfy the level-matching condition.
In the toroidal membrane case, we show that the Faddeev-Popov ghost and
anti-ghost corresponding to the global constraints are essentially
free even if we adopt any gauge fixing condition for the local
constraint. We discuss the quantum double-dimensional reduction of the
wrapped supermembrane with the global constraints.
\end{abstract}

%%%%%%%%%%%%%%%%%%%%%%%%%%%%%%%%%%
\section{Introduction}
%%%%%%%%%%%%%%%%%%%%%%%%%%%%%%%%%%
Light-cone gauge formalism is useful in string/M-theory.\footnote{In
this paper, we use a convention of the light-cone coordinates as
$x^\mu=(x^{+},x^{-},x^i)$ where $x^{\pm}=(x^0 \pm x^{D-1})/\sqrt{2}$
and the transverse coordinates are $x^i$ $(i=1,2,\cdots,(D-2))$. }
In M-theory, we know the light-cone gauge formalism \cite{BFSS},
while the covariant formalism is not yet known.
In the light-cone gauge, string/M-theory is formulated with only
the transverse degrees of freedom, however, all of them are not
independent but are subject to the constraints. For example, there
exists the level-matching condition in the closed string theory.
Such constraints originate from the residual symmetries in the
light-cone gauge: The residual symmetries are the length-preserving
and the area-preserving diffeomorphisms in the closed string and
membrane theory, respectively.

On the membrane with the non-trivial space-sheet topology, we can
decompose the generators into the co-exact and harmonic parts, which
correspond to the local and global constraints for the transverse
degrees of freedom, respectively \cite{FI,BPS,dWMN}.\footnote{In the
length-preserving diffeomorphisms of the closed string, there is no
local constraint, though the global constraint exists.}
The co-exact part generates an invariant subgroup of the group of
the area-preserving diffeomorphisms \cite{dWMN}.
The subgroup can be viewed as the $N\to\infty$ limit
of $U(N)$ and hence it can be regularized by a finite dimensional
group $U(N)$ \cite{Hop,dWHN}.

It is well known that the supermembrane in eleven dimensions is
related to type-IIA superstring in ten dimensions through the
double-dimensional reduction.
It was shown classically \cite{DHIS}, however, it is not obvious whether
it holds also in quantum theory \cite{Rus,SY}.
Recently, several authors \cite{SY,UY} analyzed the double-dimensional
reduction quantum mechanically with the light-cone wrapped
supermembrane action.
In the analyses, however, the global constraints were not taken into
account, though the space-sheet topology of the wrapped supermembrane
is a torus.

The purpose of this paper is to formulate the light-cone toroidal
membrane theory as a gauge theory of the area-preserving
diffeomorphisms incorporating the global constraints.
And based on the result, we reconsider the problem of the quantum
mechanical double-dimensional reduction with the light-cone wrapped
supermembrane action.
The plan of this paper is as follows. In section \ref{secLCT},
as a warm-up, we formulate light-cone closed string theory as a gauge
theory of the length-preserving diffeomorphisms.
We fix the gauge and calculate the BRS charge to determine the
physical states.
In section \ref{s:MEM}, we formulate light-cone toroidal membrane
theory as a gauge theory of (the total group of) the area-preserving
diffeomorphisms. In this case also, we fix the gauge and calculate the
BRS charge. By using the gauge fixed action, we discuss the quantum
double-dimensional reduction of the wrapped supermembrane with the
global constraints.

%%%%%%%%%%%%%%%%%%%%%%%%%%%%%%%%%%%%%%%%%%%%%%%%%%%%%%%%%%%%%%%%%%%
\section{Gauging the light-cone closed string}\label{secLCT}
%%%%%%%%%%%%%%%%%%%%%%%%%%%%%%%%%%%%%%%%%%%%%%%%%%%%%%%%%%%%%%%%%%%
Our starting point is the following Polyakov action of a (bosonic)
closed string,
\begin{equation}
  S_0^{st}=-\frac{T_{st}}{2}\int d\tau d\sigma\,\sqrt{-g}\,
  g^{ab}\p_a X^{\mu}\p_b X^{\nu}\eta_{\mu\nu},
\end{equation}
where the indices $\mu,\nu$ run through 0,1,$\cdots$,25
and $a,b$ take $\tau,\sigma$.
Adopting the conformal gauge of the world-sheet metric,
we obtain the following action and constrains,
\begin{eqnarray}
&&S_{cg}^{st}=\frac{T_{st}}{2}\int d\tau d\sigma\left[
	(\ptau X^{\mu})^2-(\psig X^{\mu})^2
	\right]\label{action},\\
&&(\ptau X^{\mu})^2 + ( \psig X^{\mu})^2=0\label{Cons.2},\\
&&\ptau X_{\mu} \psig X^{\mu}=0\label{Cons.1},
\end{eqnarray}
and the canonical momentum is given by $P^{\mu}=T_{st}\ptau X^{\mu}$.
As is well known, there is a residual gauge symmetry in the action
(\ref{action}). To fix the residual gauge symmetry,
we adopt the light-cone gauge,
\begin{equation}
  X^{+}=x^+ + \frac{p^+}{2\pi T_{st}}\,\tau,
\end{equation}
where $p^{+}$ is the total light-cone momentum
defined by $p^{+}=\int_0^{2\pi}d\sigma P^{+}$.
In the light-cone gauge, the action (\ref{action}) and the constraints
(\ref{Cons.2}) and (\ref{Cons.1}) are rewritten by
\begin{eqnarray}
  S_{lc}^{st}&=&\frac{T_{st}}{2}\int d\tau \int_0^{2\pi}d\sigma\left[
	(\ptau X^i)^2-(\psig X^i)^2\right]\label{LCaction},\\
  \ptau X^{-}&=&\frac{\pi T_{st}}{p^+}\left[(\ptau X^i)^2
	 + ( \psig X^{i})^2\right],\label{Cons.2'}\\
  \psig X^{-}&=&\frac{2\pi T_{st}}{p^+}\ptau X^i
	\psig X^i.\label{Cons.1'}
\end{eqnarray}
Eq.(\ref{Cons.2'}) stands for the Hamiltonian $P^-=T_{st}\ptau
X^-$ which is derived from the action (\ref{LCaction}).
Eq.(\ref{Cons.1'}) can be solved for $X^-$ in terms of the transverse
coordinates $X^i$ (with an undetermined integration constant).
In the light-cone gauge, we can describe the system only by the
transverse coordinates $X^i$, however, all the transverse degrees of
freedom are not always independent. Actually, by integrating over
$\sigma$ on both sides of eq.(\ref{Cons.1'}), we obtain the following
global constraint,
\begin{eqnarray}
  \int_0^{2\pi} d\sigma (\ptau X^i\psig X^i)=0.\label{global}
\end{eqnarray}
Hence the light-cone closed string theory is described by the action
(\ref{LCaction}) with the global constraint (\ref{global}).

Now we incorporate the global constraint (\ref{global}) as a  gauge
symmetry. First, note that the action (\ref{LCaction}) is invariant
under a constant shift of $\sigma$ coordinate,
\begin{equation}
  \delta X^i=-\epsilon_0\,\psig X^i,
\end{equation}
where the parameter $\epsilon_0$ is independent of both $\tau$ and
$\sigma$ coordinates, and hence this is a global transformation.
Next, we extend the global invariance of the action (\ref{LCaction})
into a local one with respect to $\tau$ coordinate by introducing a
gauge degree of freedom. We have the following gauge theory,
\begin{eqnarray}
  S_g^{st}&=&\frac{T_{st}}{2}\int d\tau \int_0^{2\pi}d\sigma\left[
     (D_{\tau} X^i)^2-(\psig X^i)^2\right],\label{LCgauge}\\
  &&D_{\tau}X^i=\ptau X^i +u\,\psig X^i,\nonumber
\end{eqnarray}
where the gauge degree of freedom $u=u(\tau)$ depends only on $\tau$
coordinate and is independent of $\sigma$ coordinate. The gauge theory
(\ref{LCgauge}) is invariant under the gauge transformations,
\begin{eqnarray}
  \delta X^i&=&-\epsilon\,\psig X^i,\label{gtrans1}\\
  \delta u &=&\ptau\epsilon.\label{gtrans2}
\end{eqnarray}
Note that the parameter $\epsilon=\epsilon(\tau)$ also depends only on
$\tau$ coordinate and is independent of $\sigma$ coordinate.
The transformations generate (time-dependent) length-preserving
diffeomorphisms.
It is easy to show the equivalence between the gauge theory
(\ref{LCgauge}) and the mechanical system of the action (\ref{LCaction})
with the global constraint (\ref{global}): In fact, the Euler-Lagrange
equations for $X^i$ and $u$ derived from $S_g^{st}$ (\ref{LCgauge})
are given by
\begin{eqnarray}
  D_{\tau}^2X^i-\psig^2 X^i&=&0,\label{EoM1}\\
  \int_0^{2\pi}d\sigma (D_{\tau}X^i\psig X^i)&=&0\label{EoM2}.
\end{eqnarray}
If we choose $u=0$ as a gauge fixing condition, eqs.(\ref{EoM1}) and
(\ref{EoM2}) agree with the Euler-Lagrange equation for $X^i$ derived
from the action (\ref{LCaction}) and the global constraint
(\ref{global}), respectively.
Thus we have shown the equivalence between the gauge theory
(\ref{LCgauge}) and the mechanical system of the action
(\ref{LCaction}) with the global constraint (\ref{global}), i.e.,
the light-cone closed string theory.

We should notice that the above equivalence is classical.
Our next task is to discuss the quantum-mechanical equivalence.
We adopt  $u+\xi B/2~(\xi:$ gauge parameter) as a gauge fixing
function. Then, according to a standard procedure of gauge fixing
\cite{KU}, the gauge fixed action of the gauge theory (\ref{LCgauge})
is given by
\begin{equation}
 \tilde{S}_g^{st}=\int d\tau \int_0^{2\pi}d\sigma\left[
    \frac{1}{2}(D_{\tau} X^i)^2-\frac{1}{2}(\psig X^i)^2
    +i\bar{c}\,\ptau c + Bu +\frac{\xi}{2}B^2\right],\label{LCgauge2}
\end{equation}
where the Faddeev-Popov (FP) ghost $c=c(\tau)$, anti-ghost
$\bar{c}=\bar{c}(\tau)$ and  Nakanishi-Lautrap (NL)
B-field\footnote{Precisely speaking, it should be refereed to as NL
B-degree of freedom since it is not a field.} $B=B(\tau)$ depend only
on $\tau$ coordinate and they are independent of $\sigma$ coordinate.
In the above gauge fixed action, we have set $T_{st}=1$ for brevity.
The action $(\ref{LCgauge2})$ is invariant under the BRS
transformations,
\begin{eqnarray}
  \brs \,X^i &=& -c\,\psig X^i,\\
  \brs \,u&=& \ptau c,\\
  \brs \,c&=&0,\\
  \brs \,\bar{c}&=&iB,\\
  \brs \,B &=&0,
\end{eqnarray}
and the BRS charge is given by
\begin{equation}
 Q_B=-\int_0^{2\pi}d\sigma (D_{\tau}X^i\psig X^i)\,c.\label{BRS}
\end{equation}

By using Dirac's method for constraint systems, it is straightforward
to quantize the action $(\ref{LCgauge2})$ in the operator formalism.
Actually, we obtain the following canonical (anti-)
commutation relations,
\begin{eqnarray}
  [\,X^i(\tau,\sigma),P^j(\tau,\sigma')\,]
  &=&i\delta^{ij}\delta(\sigma-\sigma'),\label{CR1}\\
  {}[\,u(\tau), X^i(\tau,\sigma)\,] &=& -i\frac{\xi}{2\pi}\psig
	 X^i(\tau,\sigma),\label{CR2}\\
  {}[\,u(\tau), P^i(\tau,\sigma)\,] &=& -i\frac{\xi}{2\pi}\psig
	 P^i(\tau,\sigma),\label{CR3}\\
  {}[\,B(\tau), X^i(\tau,\sigma)\,] &=& \frac{i}{2\pi}\psig
	X^i(\tau,\sigma),\label{CR4}\\
  {}[\,B(\tau), P^i(\tau,\sigma)\,] &=& \frac{i}{2\pi}\psig
	P^i(\tau,\sigma),\label{CR5}\\
  \{\,c(\tau),\bar{c}(\tau)\,\}&=&\frac{1}{2\pi},\label{CR6}
\end{eqnarray}
where $P^i=D_{\tau}X^i$ is the canonical momentum of $X^i$.
Furthermore, we give the Euler-Lagrange equations,
\begin{eqnarray}
  \frac{\delta \tilde{S}_g^{st}}{\delta X^i}=0
    &\To&D_{\tau}^2 X^i-\psig^2 X^i=0,\label{EL1}\\
  \frac{\delta \tilde{S}_g^{st}}{\delta u}=0
    &\To&\frac{1}{2\pi}\int_0^{2\pi}d\sigma
	 (D_{\tau}X^i\psig X^i)+B=0,\label{EL2}\\
  \frac{\delta \tilde{S}_g^{st}}{\delta B}=0
    &\To&u+\xi B=0,\label{EL3}\\
  \frac{\delta \tilde{S}_g^{st}}{\delta c}
    =\frac{\delta \tilde{S}_g^{st}}{\delta \bar{c}}=0
    &\To&\ptau \bar{c}= \ptau c=0.\label{EL4}
\end{eqnarray}
The above Euler-Lagrange equations are, of course,
consistent with the Heisenberg equations for the action
(\ref{LCgauge2}) obtained by using the canonical commutation relations
(\ref{CR1})-(\ref{CR6}) and the Hamiltonian.

Next, we solve the Euler-Lagrange equations (\ref{EL1})-(\ref{EL4}).
Henceforth, we choose the Landau gauge ($\xi=0$) only for simplicity.
In the Landau gauge, we have $u=0$ from eq.(\ref{EL3}),
and hence $D_{\tau}=\ptau$.  Then we can solve eq.(\ref{EL1}) in
terms of the oscillation modes,
\begin{equation}
  X^i(\tau,\sigma)=x^i+\frac{p^i}{2\pi}\,\tau +\frac{i}{2\sqrt{\pi}}
	\sum_{n\ne 0}\left(\frac{1}{n}\alpha_n^i\,e^{-in(\tau-\sigma)}
	 +\frac{1}{n}\tilde{\alpha}_n^i\,
	e^{-in(\tau+\sigma)}\right),\label{mode}
\end{equation}
where use has been made of the boundary condition of the closed string,
$X^i(\tau,\sigma)=X^i(\tau,\sigma +2\pi)$.
The canonical momentum $P^i$ is given by
\begin{equation}
  P^i(\tau,\sigma)=\ptau X^i (\tau,\sigma)
	=\frac{p^i}{2\pi} +\frac{1}{2\sqrt{\pi}}
	\sum_{n\ne 0}\left(\alpha_n^i\, e^{-in(\tau-\sigma)}
	+\tilde{\alpha}_n^i\,e^{-in(\tau+\sigma)}\right).
\end{equation}
Eq.(\ref{CR1}) and the above solutions lead to the commutation
relations for the oscillation modes,
\begin{equation}
  [x^i,p^j]=i\delta^{ij},\quad[\alpha_n^i,\alpha_m^j]=
    [\tilde{\alpha}_n^i,\tilde{\alpha}_m^j]=n\,\delta_{n+m,0}\,\delta^{ij}.
\end{equation}
Then we can construct the Fock space for the $X^i$ sector.
Furthermore, from eqs.(\ref{EL2}) and (\ref{mode}), we can solve for
$B$ in terms of the oscillation modes,
\begin{equation}
  B(\tau)=-\frac{1}{2\pi}\int_0^{2\pi}d\sigma\,\ptau X^i\psig X^i\nn
  =\frac{1}{2\pi}(N-\tilde{N}),
\end{equation}
where $N=\sum_{n > 0}\alpha_{-n}^i\alpha_n^i$ and $\tilde{N}
=\sum_{n>0}\tilde{\alpha}_{-n}^i\tilde{\alpha}_n^i$.
As for the FP ghost sector, from eqs.(\ref{EL4}) we can solve for $c$
and $\bar{c}$,
\begin{equation}
  c(\tau)=c_0,\quad\bar{c}(\tau)=\bar{c}_0.\label{ghost}
\end{equation}
Since the canonical anti-commutation relation is given by
\begin{equation}
  \{c_0,\bar{c}_0\}=\frac{1}{2\pi},
\end{equation}
the Fock space for the FP ghost sector is spanned by two states,
$\ket{\!\uparrow\,}$ and $\ket{\!\downarrow\,}$, which satisfy
\begin{eqnarray}
  c_0 \ket{\!\downarrow\,}= \ket{\!\uparrow\,},\quad
  \bar{c}_0 \ket{\!\uparrow\,}= \ket{\!\downarrow\,}.
\end{eqnarray}
Finally, we define the physical states according to the standard
procedure \cite{KO}. Substituting eqs.(\ref{mode}) and (\ref{ghost})
into the BRS charge (\ref{BRS}), we obtain
\begin{equation}
  Q_B=(N-\tilde{N})\,c_0.
\end{equation}
We adopt the state $\ket{\!\downarrow\,}$ as a physical state of
the FP ghost sector.\footnote{A similar degeneracy of the two
zero-mode states in the FP ghost sector appears in the covariant
quantization of a bosonic string based on the BRS invariance
\cite{KO2}.} Then the physical state condition,
\begin{equation}
  Q_B \ket{phys}\otimes\ket{\!\downarrow\,}=
    (N-\tilde{N}) \ket{phys}\otimes\ket{\!\uparrow\,}=0,
\end{equation}
leads to the condition that the physical states of the $X^i$ sector
($|phys \rangle $) satisfy the level-matching condition
($N-\tilde{N} =0$).
Thus, we have shown that the physical states in the gauge theory
(\ref{LCgauge}) agree with the states in the light-cone closed
string theory of the action (\ref{LCaction}) with the global
constraint (\ref{global}), i.e., both theories are
equivalent quantum mechanically.

%%%%%%%%%%%%%%%%%%%%%%%%%%%%%%%%%%%%%%%%%%%%%%%%%%%%%%%%%%%%%%%%%%%
\section{Gauging the light-cone toroidal membrane}\label{s:MEM}
%%%%%%%%%%%%%%%%%%%%%%%%%%%%%%%%%%%%%%%%%%%%%%%%%%%%%%%%%%%%%%%%%%%
In this section we start off with the following Polyakov-type action
of a (bosonic) membrane,
\begin{equation}
  S_0=-\frac{T}{2}\int d\tau d\sigma d\rho
  \sqrt{-g}\,(g^{ab}\p_{a} X^{\mu}\p_{b} X^{\nu}\,\eta_{\mu\nu}-1),
\end{equation}
where the indices $\mu,\nu$ run through $0,1,\cdots,D-1$ and $a,b$
take $\tau,\sigma$ and $\rho$.
We adopt the following gauge for the world-volume metric,
\begin{eqnarray}
  g_{ab}= \left(\begin{array}{cccc}
	-L^{-2}h&\bf0&\\
	\bf0 &h_{\alpha\beta}&
	\end{array}\right),
\end{eqnarray}
where $h_{\alpha\beta}~(\alpha,\beta=\sigma,\rho)$ is a metric on the
space-sheet of the membrane, $h=\det h_{\alpha\beta}$ and $L$ is an
arbitrary length parameter.\footnote{Note that the mass dimensions of
the parameters, $\tau,\sigma$ and $\rho$, are $0$ and that of the
world-volume metric $g_{ab}$ is $-2$.}
In this gauge the action and the constraints are given by
\begin{eqnarray}
  &&S_{cg}=\frac{LT}{2}\int d\tau d\sigma d\rho \left[(\ptau X^{\mu})^2
        -\frac{1}{2L^2}\{X^{\mu},X^{\nu}\}^2\right]\label{Maction},\\
  &&(\ptau X^{\mu})^2 +
	\frac{1}{2L^2}\{X^{\mu},X^{\nu}\}^2=0\label{MCons.2},\\
  &&\ptau X_{\mu}\,\p_{\alpha} X^{\mu}=0,\label{MCons.1}
\end{eqnarray}
where the Poisson bracket is defined by
\begin{equation}
  \{X^{\mu},X^{\nu}\}\equiv\epsilon^{\alpha\beta}\p_{\alpha}
	 X^{\mu}\p_{\beta} X^{\nu}=\psig X^{\mu}\prho X^{\nu}
	-\prho X^{\mu}\psig X^{\nu}.
\end{equation}
The canonical momentum is given by $P^{\mu}=LT\ptau X^{\mu}$.
In this case also, there is a residual symmetry in $S_{cg}$
(\ref{Maction}) and in order to fix it we adopt the light-cone gauge,
\begin{equation}
  X^{+}=x^+ + \frac{p^+}{(2\pi)^2 LT}\,\tau,
\end{equation}
where $p^{+}$ is the total light-cone momentum defined by
$p^{+}=\int_0^{2\pi}d\sigma d\rho\,P^{+}$.
In the light-cone gauge, the action (\ref{Maction}) and the constraints
(\ref{MCons.2}) and (\ref{MCons.1}), respectively, are rewritten by
\begin{eqnarray}
  S_{lc}&=&\frac{LT}{2}\int d\tau\int_0^{2\pi}d\sigma d\rho\left[
	(\ptau X^i)^2-\frac{1}{2L^2}\{X^{i},X^{j}\}^2\right]
	\label{MLCaction},\\
  \ptau X^{-}&=&\frac{(2\pi)^2 LT}{2p^+}\left[(\ptau X^i)^2
	+ \frac{1}{2L^2}\{X^{i},X^{j}\}^2\right]\label{MCons.2'},\\
  \p_{\alpha}X^{-}&=&\frac{(2\pi)^2 LT}{p^+} \ptau X^i\p_{\alpha}
	X^i\label{MCons.1'}.
\end{eqnarray}
In this gauge eq.(\ref{MCons.2'}) stands for the Hamiltonian,
$P^-=LT\ptau X^-$, for the action (\ref{MLCaction}).
Eq.(\ref{MCons.1'}) can be solved for $X^-$ in terms of the transverse
coordinates $X^i$ (with an undetermined integration constant). Thus,
in the light-cone gauge, the system is described only with the
transverse coordinates $X^i$. However, all the transverse degrees of
freedom are not always independent.
The integrability condition for eq.(\ref{MCons.1'}),
$\epsilon^{\alpha\beta}\p_\alpha\p_\beta X^-=0$, leads to
\begin{equation}
  \Phi_0(\sigma,\rho)\equiv\{\ptau X^i,X^i\}=0.\label{local}
\end{equation}
This is locally equivalent to eq.(\ref{MCons.1'}) and we call it the
local constraint.\footnote{As for the closed string in the previous
section, there is no counterpart of the local constraint.}
As for the membrane with the non-trivial space-sheet topology,
we must further impose the global constraints.
Henceforth we consider the toroidal membrane for definiteness.
Then, $X^-$ is a periodic function with respect to both $\sigma$ and
$\rho$ coordinates. Integrating both sides of
eq.(\ref{MCons.1'}) over either $\sigma$ or $\rho$  for
$\alpha=\sigma$ or $\alpha=\rho$, respectively, we obtain the
following global constraints,
\begin{eqnarray}
  \Phi_1(\rho)&\equiv&\int_0^{2\pi} d\sigma (\ptau X^i\psig X^i)
	=0,\label{Mglobal1}\\
  \Phi_2(\sigma)&\equiv&\int_0^{2\pi} d\rho (\ptau X^i\prho X^i)
	=0.\label{Mglobal2}
\end{eqnarray}
Such global constraints (\ref{Mglobal1}) and (\ref{Mglobal2}),
however, are not completely independent of the local constraint
(\ref{local}).
In fact, by integrating (\ref{local}) over $\sigma$, we get
\begin{eqnarray}
  0= -\int_0^{2\pi}d\sigma\, \Phi_0(\sigma,\rho)
    =\, \prho \int_0^{2\pi}d\sigma(\ptau X^i\psig X^i)
    =\prho\,\Phi_1(\rho)\,.
\end{eqnarray}
This means that the constraint (\ref{Mglobal1}) is already included in
the local constraint (\ref{local}) except for the $\rho$-independent
mode of $\Phi_1(\rho)$. Thus, in addition to the local constraint
(\ref{local}) we may impose the following global constraint instead
of eq.(\ref{Mglobal1}),
\begin{equation}
 \Phi_\sigma\equiv\int_0^{2\pi} d\sigma d\rho
	(\ptau X^i\psig X^i)=0.\label{Mglobal'1}
\end{equation}
Similarly, instead of eq.(\ref{Mglobal2}), we may impose the following
global constraint,
\begin{equation}
  \Phi_\rho\equiv\int_0^{2\pi} d\sigma d
	\rho(\ptau X^i\prho X^i)=0.\label{Mglobal'2}
\end{equation}
Thus, the light-cone toroidal membrane theory is described by the
action (\ref{MLCaction}) with the local constraint (\ref{local})
and the global constraints (\ref{Mglobal'1}) and (\ref{Mglobal'2}).

Now we incorporate the local constraint (\ref{local}) and the global
constraints (\ref{Mglobal'1}) and (\ref{Mglobal'2}) as a gauge
theory. Actually, the following gauge theory is equivalent to the
system described by the action (\ref{MLCaction}) with the constraints
(\ref{local}), (\ref{Mglobal'1}) and (\ref{Mglobal'2}),
\begin{eqnarray}
  S_g&=&\frac{LT}{2}\int d\tau \int_0^{2\pi}d\sigma d\rho\left[
   (D_{\tau}X^i)^2-\frac{1}{2L^2}\{X^{i},X^{j}\}^2\right]\label{MLCgauge},\\
  &&D_{\tau}X^i=\ptau X^i+u^{\alpha}\p_{\alpha}X^i\nonumber,
\end{eqnarray}
where $u^{\alpha}$ is a gauge field satisfying the condition,
$\p_{\alpha}u^{\alpha}=0$.
The action (\ref{MLCgauge}) is invariant under the gauge
transformations,
\begin{eqnarray}
 \delta X^i&=&-\epsilon^{\alpha} \p_{\alpha} X^i,\label{Mgtrans1}\\
 \delta u^{\alpha} &=&\ptau\epsilon^{\alpha}
	+\p_{\beta} \epsilon^{\alpha} u^{\beta}
	-\p_{\beta} u^{\alpha} \epsilon^{\beta},\label{Mgtrans2}
\end{eqnarray}
where the parameter $\epsilon^{\alpha}$  also satisfies the condition,
$\p_{\alpha}\epsilon^{\alpha}=0$.
The transformations generate (time-dependent) area-preserving
diffeomorphisms. It is not difficult to see the equivalence between
the gauge theory (\ref{MLCgauge}) and the mechanical system of the
action (\ref{MLCaction}) with the constraints, (\ref{local}),
(\ref{Mglobal'1}) and (\ref{Mglobal'2}).
First, due to $\p_{\alpha}u^{\alpha}=0$, we can decompose the gauge
field $u^{\alpha}$ into the following co-exact and harmonic parts
\cite{FI,BPS,dWMN},
\begin{equation}
  u^{\alpha}=\frac{1}{L}\,\epsilon^{\alpha\beta}\,\p_{\beta}A
	+\frac{1}{L}\,a^{\alpha},\label{gauge}
\end{equation}
where the co-exact part $A=A(\tau,\sigma,\rho)$ is an
arbitrary periodic function with respect to $\sigma$ and $\rho$
coordinates and the harmonic part $a^{\alpha}=a^{\alpha}(\tau)$
is independent of both $\sigma$ and $\rho$ coordinates.
Similarly, due to $\p_{\alpha}\epsilon^{\alpha}=0$, we can decompose
the parameter $\epsilon^{\alpha}$ into the following co-exact and
harmonic parts,
\begin{equation}\label{parameter}
  \epsilon^{\alpha}=\frac{1}{L}\,\epsilon^{\alpha\beta}\,
	\p_{\beta}\Lambda +\frac{1}{L}\,\lambda^{\alpha},
\end{equation}
where the co-exact part $\Lambda=\Lambda(\tau,\sigma,\rho)$ is an
arbitrary periodic function with respect to $\sigma$ and $\rho$
coordinates and the harmonic part
$\lambda^{\alpha}=\lambda^{\alpha}(\tau)$ depends only on $\tau$.
The co-exact part $\Lambda$ generates an invariant subgroup
of the total group of the area-preserving diffeomorphisms \cite{dWMN}.
Using the expressions (\ref{gauge}) and (\ref{parameter}),
we can decompose the gauge transformations (\ref{Mgtrans1})
and (\ref{Mgtrans2}) as
\begin{eqnarray}
  \delta X^i&=&\frac{1}{L}\,\{\Lambda,X^i\}-\frac{1}{L}\,
	\lambda^{\alpha}\p_{\alpha}X^i,\label{Mgtrans1'}\\
  \delta A~ &=&\ptau \Lambda +\frac{1}{L}\,\{\Lambda,A\}
    +\frac{1}{L}\,\p_{\alpha}\Lambda\, a^{\alpha}
    -\frac{1}{L}\,\lambda^{\alpha}\p_{\alpha}A,\label{Mgtrans2'}\\
  \delta a^{\alpha}&=&\ptau\lambda^{\alpha}.\label{Mgtrans3'}
\end{eqnarray}
Furthermore, by using the expression (\ref{gauge}),
the covariant derivative in eq.(\ref{MLCgauge}) becomes
\begin{equation}
  D_{\tau}X^i=\ptau X^i -\frac{1}{L}\{A,X^i\}
	+\frac{1}{L}a^{\alpha}\p_{\alpha}X^i.\label{CD2}
\end{equation}
The Euler-Lagrange equations for $X^i, A$ and $a^{\alpha}$ derived
from the action (\ref{MLCgauge}), respectively, are given by
\begin{eqnarray}
  &&D_{\tau}^2X^i - \frac{1}{L^2}\{\{X^i,X^j\},X^j\}=0,\label{MEOM1}\\
  &&\{D_{\tau} X^i,X^i\}=0,\label{MEOM2}\\
  &&\int_0^{2\pi} d\sigma d\rho\,(D_{\tau} X^i\psig X^i)=\int_0^{2\pi}
	d\sigma d\rho\,(D_{\tau} X^i\prho X^i)=0.\label{MEOM3}
\end{eqnarray}
If we choose  $A=a^{\alpha}=0$ as gauge fixing conditions for the
gauge transformations (\ref{Mgtrans1'})-(\ref{Mgtrans3'}),
eqs.(\ref{MEOM1}), (\ref{MEOM2}) and (\ref{MEOM3}) agree with the
Euler-Lagrange equation for $X^i$ derived from the action
(\ref{MLCaction}), the local constraint (\ref{local}) and
the global constraints (\ref{Mglobal'1}) and (\ref{Mglobal'2}),
respectively. Thus it has been shown that the gauge theory
(\ref{MLCgauge}) is equivalent to the mechanical system described by
the action (\ref{MLCaction}) with the constraints (\ref{local}),
(\ref{Mglobal'1}) and (\ref{Mglobal'2}), i.e., the light-cone
toroidal membrane theory.

A definitive difference between the closed string theory in the
previous section and the toroidal membrane theory in this section is
that it is easy to quantize the action (\ref{LCaction}) or
(\ref{LCgauge}) of the closed string, while it is not for the action
(\ref{MLCaction}) or (\ref{MLCgauge}) of the toroidal membrane due to
non-linearity and non-renormalizability.
As for the gauge theory, however, which is obtained by restricting
the group of the area-preserving diffeomorphisms on the invariant
subgroup generated by the parameter $\Lambda$, we can adopt the matrix
regularization, so that we can quantize the gauge theory
perturbatively.
The action of such a gauge theory is as follows,\footnote{The action
corresponds to that of a spherical membrane, which has no harmonic
part from the outset.}
\begin{equation}
  S_{g'}=\frac{LT}{2}\int d\tau \int_0^{2\pi}d\sigma d\rho\left[
    \left(\ptau X^i -\frac{1}{L}\{A,X^i\}\right)^2
	-\frac{1}{2L^2}\{X^{i},X^{j}\}^2\right]\label{MLCgaugeSp},
\end{equation}
and the gauge transformations are given by
\begin{eqnarray}
\delta X^i&=&\frac{1}{L}\{\Lambda,X^i\},\label{Mgtrans1'Sp}\\
\delta A&=&\ptau\Lambda +\frac{1}{L}\{\Lambda,A\}.\label{Mgtrans2'Sp}
\end{eqnarray}
The matrix regularization is to regularize the infinite dimensional
subgroup of the total group of the area-preserving diffeomorphisms
by a finite dimensional group $U(N)$ \cite{Hop,dWHN}.
In the matrix regularization, the infinite degrees of freedom due to
the dependence of $\sigma$ and $\rho$ coordinates are mapped to
$N\times N$ finite ones of hermitian matrices and the double integrals
over $\sigma$ and $\rho$ and the Poisson bracket are mapped as
follows,
\begin{eqnarray}
  \int_0^{2\pi}d\sigma d\rho &\to& \mbox{Tr}\,,\label{tr}\\
  \{\,\cdot\,,\,\cdot\,\} &\to&
	  -i\,[\,\cdot\,,\,\cdot\,]\,.\label{com}
\end{eqnarray}
We consider the matrix regularization of the gauge theory
(\ref{MLCgauge}), which is invariant under the gauge transformation
including the parameter $\lambda^{\alpha}$.
The simple maps, (\ref{tr}) and (\ref{com}), however, cannot be
applied straightforwardly because the covariant derivative (\ref{CD2})
includes naked derivatives of $\psig$ and $\prho$ which are not
described by the Poisson brackets.

Let us fix the gauge of the action $S_g$ in eq.(\ref{MLCgauge}).
Henceforth we set $LT=1$ for brevity. Following the standard procedure
\cite{KU} we get the gauge fixed action,
\begin{eqnarray}
  \tilde{S}_g&=& S_g +  \int d\tau \int_0^{2\pi}d\sigma d\rho
	\, \brs(-i\bar{c}^\alpha a^\alpha
	-i\bar{C}(\ptau A+\frac{\xi}{2}B) ) \nn
  &=&\int d\tau \int_0^{2\pi}d\sigma d\rho\Biggl[
    \frac{1}{2}(D_{\tau} X^i)^2-\frac{1}{4L^2}\{X^{i},X^{j}\}^2
    +i\bar{c}^{\alpha}\ptau c^{\alpha} +b^{\alpha}a^{\alpha}\nn
  && \hspace{5ex}- i\ptau\bar{C}\left(\ptau C +\frac{1}{L}\{C,A\}
    +\frac{1}{L}\p_{\alpha}C a^{\alpha}
    -\frac{1}{L}c^{\alpha}\p_{\alpha}A\right)+ B\ptau A
	+\frac{\xi}{2}B^2 \Biggr],\label{MLCgauge2}
\end{eqnarray}
where $\xi$ is a gauge parameter, $C(\tau,\sigma,\rho)$,
$\bar{C}(\tau,\sigma,\rho)$ and $B(\tau,\sigma,\rho)$ are the FP
ghost, anti-ghost and NL B-field associated with the co-exact
part of the gauge transformation, while the FP ghost
$c^{\alpha}(\tau)$, anti-ghost $\bar{c}^{\alpha}(\tau)$ and
NL B-field $b^{\alpha}(\tau)$ are associated with the harmonic part
$\lambda^{\alpha}$ and they depend only on $\tau$ coordinate.
$\tilde{S}_g$ is invariant under the BRS transformations,
\begin{eqnarray}
  \brs\,X^i &=& \frac{1}{L}\{ C, X^i\}-\frac{1}{L}
	 c^{\alpha}\p_{\alpha} X^i,\\
  \brs\,A&=& \ptau C +\frac{1}{L}\{ C, A\}
	+\frac{1}{L}\p_{\alpha}C a^{\alpha}-\frac{1}{L}
	 c^{\alpha}\p_{\alpha} A,\\
  \brs\,a^{\alpha}&=&\ptau c^{\alpha},\\
  \brs\,C&=&\frac{1}{2L}\{C,C\}
	-\frac{1}{L} c^{\alpha} \p_{\alpha}C,\\
  \brs\,c^{\alpha}&=&0,\\
  \brs\,\bar{C}&=&i B,\\
  \brs\,\bar{c}^{\alpha}&=&i b^{\alpha},\\
  \brs\,B &=& \brs\,b^{\alpha} =\,0.
\end{eqnarray}
The BRS charge is given by
\begin{eqnarray}
  Q_B&=&-\frac{1}{L}\int_0^{2\pi} d\sigma d\rho \biggl[
    \{D_{\tau}X^i,X^i\}\,C + D_{\tau}X^i\p_{\alpha}X^i\, c^{\alpha}\nn
  &&\hspace{15ex} -\frac{i}{2}\{\ptau\bar{C},C\}\,C
    -i\ptau\bar{C}\,\p_{\alpha}C\,c^{\alpha}\biggr].
\end{eqnarray}
We concentrate on the contribution of the BRS quartet
$\{a^\alpha,b^\alpha,c^\alpha,\bar{c}^\alpha\}$ associated
with the global constraints, which we call GC-quartet.
We extract the terms which include the member(s) of the GC-quartet
from $\tilde{S}_g$ (\ref{MLCgauge2}),
\begin{eqnarray}
  S_{GCq}&=&\int d\tau \int_0^{2\pi}d\sigma d\rho\Biggl[
    \left(\ptau X^i -\frac{1}{L}\{A,X^i\}\right)
	\frac{1}{L}a^\alpha \p_\alpha X^i
    + \frac{1}{2L^2}\left(a^\alpha \p_\alpha X^i\right)^2\nn
   && \hspace{17ex}  +i\bar{c}^{\alpha}\ptau c^{\alpha}
   +b^{\alpha}a^{\alpha} - i\frac{1}{L}\ptau\bar{C}\left(
    \p_{\alpha}C\,a^{\alpha}-c^{\alpha}\p_{\alpha}A\right)\Biggr]\nn
  &=&\int d\tau \int_0^{2\pi}d\sigma d\rho\left[
    \tilde{b}^\alpha a^\alpha + i\bar{c}^{\alpha}\ptau c^{\alpha}
    +i\frac{1}{L}\ptau\bar{C}\left(c^{\alpha}\p_{\alpha}A\right)\right],
    \label{GCq}
\end{eqnarray}
where
\begin{equation}
  \tilde{b}^\alpha=b^\alpha + \int^{2\pi}_0
    \frac{d\sigma d\rho}{(2\pi)^2} \frac{1}{L}\left[
    \left(\ptau X^i-\frac{1}{L}\{A,X^i\}\right)\p_\alpha X^i
    +\frac{1}{2L}\p_\alpha X^i a^\beta\p_\beta X^i
	- i\ptau\bar{C}\p_\alpha C\right].
\end{equation}
The action $S_{GCq}$ (\ref{GCq}) consists of free part (the first two
terms) and the interaction part (the last term). However, we should
notice that the interaction term never contributes to any correlation
functions of $X^i, A, C, \bar{C}$ and $B$ {\it since there is no
interaction term which includes both $\bar{c}^\alpha$ and $C$ in the
total action $\tilde{S}_g$.}\footnote{In the path-integral language,
this is due to the fact that the FP determinant of the triangular
matrix is equal to that of the diagonal matrix.}
This means that  $S_{GCq}$ is essentially free and hence it
contributes an overall trivial factor to the correlation functions
of $X^i, A, C, \bar{C}$ and $B$ in the path-integral formalism.
Thus, $\tilde{S}_g$ (\ref{MLCgauge2}) is equivalent to the gauge fixed
action  for the gauge theory $S_{g'}$ (\ref{MLCgaugeSp}) up to the
essentially free action for the GC-quartet.
Note that after (path-)integrating out the GC-quartet we can
matrix-regularize the action, i.e., $\tilde{S}_g-S_{GCq}$, by the
simple maps (\ref{tr}) and (\ref{com}).
We should also notice that the above mentioned facts always hold
if we adopt $\brs(-i\bar{c}^\alpha a^{\alpha})$ to fix the
$\lambda^{\alpha}$ gauge transformation and do not use the GC-quartet
in the gauge fixing function $F_\Lambda$ of the $\Lambda$ gauge
transformation, i.e., we adopt
$\brs{(-i\bar{c}^\alpha a^{\alpha}-i\bar{C}F_\Lambda)}$ where
$F_\Lambda$ is an arbitrary function which does not depend on any of
$\{a^\alpha,b^\alpha,c^\alpha,\bar{c}^\alpha\}$.\footnote{
We can adopt such a weaker condition for $F_\Lambda$ as $\p
F_\Lambda/\p\bar{c}^\alpha=0$.}
Hence, as to the $\Lambda$ gauge transformation, we can adopt not only
$F_\Lambda=\ptau A +\xi B/2$ but also such a gauge as was
used in the wrapped supermembrane theory \cite{SY,UY}.
 
Finally we comment on the quantum mechanical study of the
double-dimensional reduction of the eleven dimensional supermembrane
\cite{SY,UY}.
In Ref.\cite{SY}, the problem was first analyzed in the world-volume
field theory of the wrapped supermembrane in the path-integral formalism.
Then in Ref.\cite{UY}, the similar analysis was performed in the
world-volume field theory in the matrix-regularized form of the
wrapped supermembrane, i.e., matrix string theory \cite{Mot,DVV}.
In those analyses, however, the global constraints, which should
be the extended version of (\ref{Mglobal'1}) and (\ref{Mglobal'2})
to the wrapped supermembrane theory, were not taken into account.
As for Ref.\cite{UY} in particular, it is because the
matrix-regularized forms of the global constraints were not obvious.
Actually, in the standard derivation \cite{Mot,DVV} of matrix string
theory based on Seiberg and Sen's arguments \cite{Sei,Sen} and the
compactification prescription by Taylor \cite{Tay}, such
matrix-regularized global constraints do not appear naturally.
The difference between $\tilde{S}_g$ (\ref{MLCgauge2}) and the gauge
fixed version of $S_{g'}$ (\ref{MLCgaugeSp}) is only the essentially
free action for the GC-quartet, \{$a^\alpha, b^\alpha, c^{\alpha},
\bar{c}^{\alpha}$\}, which do not affect the quantum mechanical study of
the double-dimensional reduction in the path-integral formalism.
This reminds us of QED in the path-integral formalism, where
the FP ghosts can be free and contribute only to the vacuum energy.
In fact, the determinant of the FP ghosts cancel out the contribution
from the longitudinal and scalar modes of the gauge field.
However such a determinant of the FP ghosts is not important to
physical amplitudes.
Thus, our results of this paper justify the analyses of Refs.\cite{SY}
and \cite{UY}, where the global constraints were not taken into
account.\\[\baselineskip]
\noindent {\bf Acknowledgments:}
The work of SU is supported in part by the Grant-in-Aid for Scientific
Research No.13135212.

%%%%%%%%%%%%%%%%%%%%%%%%%%%%%%%%%

%%%%%%%%%%%%%%%%%%%%%%%%%%%%%%%%%
%%%%%%%%%%%%%%%%%%%%%%%%%%%%%%%%%

\begin{thebibliography}{99}
%%%%%%%%%%%%%%%%%%%%%%%%%%%%%%%%%


\bibitem{BFSS} T.~Banks, W.~Fischler, S.~H.~Shenker and L.~Susskind,
	``M theory as a matrix model: A conjecture,''
	Phys.\ Rev.\ D {\bf 55}, 5112 (1997)
	[arXiv:hep-th/9610043].
%%CITATION = HEP-TH 9610043;%%

\bibitem{FI} E.~G.~Floratos and J.~Iliopoulos,
	``A Note On The Classical Symmetries Of The Closed Bosonic
	Membranes,'' Phys.\ Lett.\ B {\bf 201}, 237 (1988).
%%CITATION = PHLTA,B201,237;%%

\bibitem{BPS} I.~Bars, C.~N.~Pope and E.~Sezgin,
	``Central Extensions Of Area Preserving Membrane Algebras,''
	Phys.\ Lett.\ B {\bf 210}, 85 (1988).
%%CITATION = PHLTA,B210,85;%%

\bibitem{dWMN} B.~de Wit, U.~Marquard and H.~Nicolai,
	``Area Preserving Diffeomorphisms And Supermembrane Lorentz
	Invariance,''
	Commun.\ Math.\ Phys.\  {\bf 128}, 39 (1990).
%%CITATION = CMPHA,128,39;%%

\bibitem{Hop} J.~Hoppe,
	``Quantum theory of a relativistic membrane,''
	M.I.T. Ph.D. thesis, (1982).

\bibitem{dWHN} B.~de Wit, J.~Hoppe and H.~Nicolai,
	``On The Quantum Mechanics Of Supermembranes,''
	Nucl.\ Phys.\ B {\bf 305}, 545 (1988).
%%CITATION = NUPHA,B305,545;%%

\bibitem{DHIS}M.~J.~Duff, P.~S.~Howe, T.~Inami and K.~S.~Stelle,
	``Superstrings In D = 10 From Supermembranes In D = 11,''
	Phys.\ Lett.\ B {\bf 191}, 70 (1987).
%%CITATION = PHLTA,B191,70;%%

\bibitem{Rus}J.~G.~Russo,
	``Supermembrane dynamics from multiple interacting strings,''
	Nucl.\ Phys.\ B {\bf 492}, 205 (1997)
	[arXiv:hep-th/9610018].
%%CITATION = HEP-TH 9610018;%%

\bibitem{SY}Y.~Sekino and T.~Yoneya,
	``From supermembrane to matrix string,''
	Nucl.\ Phys.\ B {\bf 619}, 22 (2001)
	[arXiv:hep-th/0108176].
%%CITATION = HEP-TH 0108176;%%

\bibitem{UY} S.~Uehara and S.~Yamada,
	``On the strong coupling region in quantum matrix string
	theory,''
	JHEP {\bf 0209}, 019 (2002) [arXiv:hep-th/0207209],
%%CITATION = HEP-TH 0207209;%%
	``On the quantum matrix string,'' arXiv:hep-th/0210261.
%%CITATION = HEP-TH 0210261;%%

\bibitem{KU} T.~Kugo and S.~Uehara,
	``General Procedure Of Gauge Fixing Based On BRS Invariance
	Principle,''
	Nucl.\ Phys.\ B {\bf 197}, 378 (1982).
%%CITATION = NUPHA,B197,378;%%

\bibitem{KO} T.~Kugo and S.~Ojima,
	``Manifestly Covariant Canonical Formulation Of Yang-Mills
	Field Theories: Physical State Subsidiary Conditions And
	Physical S Matrix Unitarity,''
	Phys.\ Lett.\ B {\bf 73}, 459 (1978).
%%CITATION = PHLTA,B73,459;%%

\bibitem{KO2} M.~Kato and K.~Ogawa,
	``Covariant Quantization Of String Based On BRS Invariance,''
	Nucl.\ Phys.\ B {\bf 212}, 443 (1983).
%%CITATION = NUPHA,B212,443;%%

\bibitem{Mot}L.~Motl,
	``Proposals on nonperturbative superstring interactions,''\\
	arXiv:hep-th/9701025.
%%CITATION = HEP-TH 9701025;%%

\bibitem{DVV}R.~Dijkgraaf, E.~Verlinde and H.~Verlinde,
	``Matrix string theory,''
	Nucl.\ Phys.\ B {\bf 500}, 43 (1997)
	[arXiv:hep-th/9703030].
%%CITATION = HEP-TH 9703030;%%

\bibitem{Sei} N.~Seiberg,
	``Why is the matrix model correct?,''
	Phys.\ Rev.\ Lett.\  {\bf 79}, 3577 (1997)
	[arXiv:hep-th/9710009].
%%CITATION = HEP-TH 9710009;%%

\bibitem{Sen} A.~Sen,
	``D0-branes on $T^n$ and matrix theory,''
	Adv.\ Theor.\ Math.\ Phys.\  {\bf 2}, 51 (1998)
	[arXiv:hep-th/9709220].
%%CITATION = HEP-TH 9709220;%%

\bibitem{Tay} W.~I.~Taylor,
	``D-brane field theory on compact spaces,''
	Phys.\ Lett.\ B {\bf 394}, 283 (1997)
	[arXiv:hep-th/9611042].
%%CITATION = HEP-TH 9611042;%%

%%%%%%%%%%%%%%%%%%%%%%%%%%%%%%%%%
\end{thebibliography}
\end{document}